\documentclass[referee]{aa} 
\usepackage[latin1]{inputenc}
\usepackage[T1]{fontenc}
\usepackage{graphicx}
\usepackage{txfonts}
\usepackage{upgreek}
\usepackage{verbatim}
\usepackage{natbib}
\usepackage{color}
\newcommand{\hm}{H$_{2}$}

\begin{document}
   \title{When sticking influences H$_2$ formation}

   \author{S.Cazaux
          \inst{1}
          \and
          S. Morisset
          \inst{2}
          \and
          M. Spaans
          \inst{1}
          \and
          A. Allouche
          \inst{2}
          }

\institute{Kapteyn Astronomical Institute, PO box 800, 9700AV Groningen, The Netherlands\\
\and 
Laboratoire de Physique des Interactions Ioniques et Mol\'eculaires, UMR CNRS 6633, Universit\'e de Provence, Campus de Saint-J\'er\^ome, case 242, 13397 Marseille Cedex 20, France}

   \date{Received xxx; accepted xxx}

 
  \abstract
{
}
{Interstellar dust grains, because of their catalytic properties, are crucial to the formation of \hm, the most abundant molecule in the Universe. The formation of molecular hydrogen strongly depends on the ability of H atoms to stick on dust grains. In this study we determine the sticking coefficient of H atoms \rm{chemisorbed}\rm\ on graphitic surfaces, and estimate its impact on the formation of \hm.
}
{The sticking probability of H atoms chemisorbed onto graphitic surfaces is obtained using  a mixed classical-quantum dynamics method. In this, the H atom is treated quantum-mechanically and the vibrational modes of the surface are treated classically. The implications of sticking for the formation of \hm\  are addressed by using \rm{Kinetic}\rm\ Monte Carlo simulations that follow how atoms stick, move and associate with each other on dust surfaces of different temperature.
}
{\rm{In our model}\rm, molecular hydrogen forms very efficiently for dust temperatures lower than 15~K through the involvement of physisorbed H atoms. At dust temperatures higher than 15~K and gas temperatures lower than 2000~K, \hm\ formation differs strongly if the H atoms coming from the gas phase have to cross a square barrier (usually considered in previous studies) or  a barrier obtained by DFT calculations to become chemisorbed. The product of sticking times efficiency can be increased by many orders of magnitude when realistic barriers are considered. If graphite phonons are taken into account in the dynamics calculations, then H atoms stick better on the surface at high energies, but the overall \hm\ formation efficiency is only slightly affected. Our results suggest that \hm\ formation can proceed efficiently in photon dominated regions, X-ray dominated regions, hot cores and in the early Universe when the first dust is available. }
{
}

   \keywords{ISM: dust, extinction -- ISM: molecules -- ISM: abundances}

   \maketitle
%

\section{Introduction}
Interstellar dust grains play a very important role in the chemistry
of the interstellar medium (ISM). Because of inefficient gas phase paths to
form \hm, dust grains are considered as the favored habitat to
form \hm\ molecules (\citealt{gould1963}). In the past decades, a plethora of
laboratory experiments and theoretical models have been developed in
order to understand how the most abundant molecule of the Universe forms. The sticking of H atoms on surfaces has received considerable attention since this mechanism governs the formation of H$_2$, but also other molecules that contain H atoms. The sticking of H atoms on dust grains can also be an important mechanism to cool interstellar gas (\citealt{spaans2000}). As H atoms arrive on the dust, they can be weakly bound (physisorbed) or strongly bound (chemisorbed) to the surface. \rm{The sticking of H in physisorption state has been highlighted by several experiments on different type of surfaces (\citealt{pirronello1997a}, \citeyear{pirronello1999}, \citeyear{pirronello2000},\citealt{perry2003}). Experiments performed on surfaces at higher temperatures revealed how H atoms stick in chemisorbed state (\citealt{zecho2002}, \citealt{hornekaer2006}, \citealt{mennella2006}). In addition, theoretical studies confirmed that the sticking of H atoms  varies strongly with gas and dust temperatures (\citealt{leitch1985}, \citealt{buch1989}, \citealt{sha2005}, \citealt{medina2008}, \citealt{morissetbis2008}, \citealt{cuppen2010}, \citealt{morisset2010}})\rm. To study the sticking of an atom on a surface, it is necessary to take into account the vibrational modes (phonons) of the surface (\citealt{burke1983}, \citealt{buch1989}, \citealt{morissetbis2008}, \citealt{morisset2010} ).  

In the ISM, dust grains are mainly carbonaceous particles or silicates, with various sizes, and a large fraction of the available surface for chemistry is in the form of very small grains or polyaromatic hydrocarbons (PAHs; \citealt{Weingartner2001}). PAHs have similar characteristics as graphite surfaces, as shown by different calculations (\citealt{jeloaica1999}, \citealt{sha2002}, \citealt{ferro2008}). \cite{jeloaica1999} studied the interaction between H atoms and a coronene,  whereas \cite{ferro2008}  and \cite{sha2002}  determined the H-graphite interaction. All these theoretical works obtained identical  characteristics: the H atom coming from the gas phase can physisorb (43~meV) or chemisorb (0.76~eV) on the graphite or PAH. \rm{Several experimental studies showed that H atoms can physisorb on carbonaceous (\citealt{pirronello1997a}, \citeyear{pirronello1999}, \citeyear{pirronello2000}), silicates (\citealt{pirronello1997b})  and graphitic (\citealt{perry2003}) surfaces, and  can also chemisorb on graphite surfaces  (\citealt{zecho2002}, \citealt{hornekaer2006}). These experiments revealed that adsorption of H atoms (physisorption and chemisorption) is of key importance to form H$_2$ for a wide range of dust and gas temperatures, as observed in the ISM.}\rm\ The H atom can chemisorb only on top of a C atom and a surface reconstruction is observed: the C atom goes out of the surface plane. This phenomenon is called puckering, and creates a barrier against chemisorption of 0.2~eV (\citealt{jeloaica1999}, \citealt{sha2002}, \citealt{ferro2003}, \citealt{hornekaer2006}). This barrier is important and does not allow an efficient sticking mechanism for atoms coming in at low energies. Therefore, sticking through chemisorption strongly depends on the energy of the incoming atoms. Also, the sticking is sensitive to the temperature of the dust since the surface excitation modes play a role in redistributing the excess energy of an atom.
\rm

The \hm\ formation occurs through Eley-Rideal (ER) or Langmuir-Hinshelwood (LH) mechanisms. In the ER mechanism, an H  atom coming from the gas phase collides with an H atom which is initially adsorbed (chemisorbed) on the graphite surface. Theoretical (\citealt{rougeau2006}, \citealt{ferro2008}) and experimental (\citealt{hornekaer2006}) studies show that H atoms are chemisorbed on the surface in dimer  configurations, and that two hydrogen dimers configurations are stable: the ortho-dimer and the para dimer (Figure 3, right panel). Density Functional Theory (DFT) calculations show that while the chemisorption of one H atom is associated with an important barrier, the formation of the para dimer is barrier-less and that the formation of the ortho dimer has a reduced barrier (\cite{rougeau2006}). The formation of \hm\ that involves chemisorbed atoms through the ER mechanism has been studied by Density Functional Theory (DFT) (\citealt{jeloaica1999}, \citealt{sha2002}, \citealt{ferro2003}) and dynamics calculations (\citealt{rutigliano2001}, \citealt{ree2002}, \citealt{sha2002},  \citealt{morisset2003}, \citeyear{morisset2004b}, \citealt{martinazzo2006chem}). Based on these calculations, different mechanisms have been proposed to contribute to the \hm\ formation through the ER mechanisms: the direct ER that involves isolated H atoms (monomers,  \citealt{sha2002}, \citealt{morisset2003}, \citeyear{morisset2004b}, \citealt{martinazzo2006chem}), barrier-less formation of \hm\ involving one H atom in a para-dimer configuration (\citealt{bachellerie2007}) and formation by diffusing H atoms in physisorbed states  (\citealt{bonfanti2007}).  
In the LH mechanism, the two H atoms are adsorbed (physisorbed) on the graphite surface, diffuse on the surface and collide to desorb in a hydrogen molecule.  This mechanism has also been studied (\citealt{morisset2004},  \citeyear{morisset2005},  \citealt{martinazzo2006phys}). The different mechanisms to form \hm\ on graphite surfaces are the following:\\
M1) LH mechanism: Two physisorbed H atoms encounter each other on the surface; \\
M2) direct Eley-Rideal mechanism involving \rm{H atom chemisorbed in a monomer (only one H atom chemisorbed on the cycle, fig. \ref{benzene}, right panel)}\rm;\\ 
M3) direct Eley-Rideal mechanism involving an H atom chemisorbed in a  dimer \rm{(H atom  chemisorbed in dimer position on the first cycle, fig. \ref{benzene}, right panel)\rm;\\
M4) fast diffusion of physisorbed H atoms that enter chemisorbed sites occupied by H atoms (monomer);\\
M5) fast diffusion of physisorbed H atoms that enter chemisorbed sites occupied by H atoms (dimer);\\
M6) direct Eley-Rideal mechanism involving an H atom in meta, 5 and S (second cycle) positions (\citealt{dumont2008}, see figure 3 right panel);\\
M7) ER mechanism by fast diffusing H atoms in the physisorption state (\citealt{bonfanti2007}).\\

\rm{Recent experimental studies (\citealt{lemaire2010}, \citealt{islam2007}) show that the ro-vibrationnal states of the \hm molecule formed on the surface can be detected (silicates and graphite). The ro-vibrationnal distribution of the \hm\ allows to classify the LH and the ER mechanisms to form the molecule as function of the dust temperature. Experimentally, it is not possible to identify the M1 to M7 mechanisms proposed to form \hm. The objective of our work is to classify the predominant mechanisms to form \hm\ as a function of gas and dust temperature.}\rm\ In each mechanism proposed, the sticking of the H atoms in physisorption sites and chemisorption sites (figure 3 left panel) is crucial to allow the \hm\ formation. In this study, we concentrate on understanding how the first H atom chemisorbs on the grain with a direct ER mechanism and how the sticking probability impacts the \hm\ formation. During the collision between an atom and a surface, the collisional energy is distributed between the vibrational excitation of the newly formed bond and the vibrational modes of the surface.Therefore, to calculate the sticking of atoms on a surface in a realistic way, one has to consider the lattice dynamics. For this purpose, we study the interaction of H atoms with graphitic surfaces, which are surfaces having similar properties as PAH surfaces. Our attention focusses particularly on these surfaces because PAHs represent a large fraction of the total surface area of dust grains (up to 50$\%$, \citealt{Weingartner2001}).

In the first section, the time dependent dynamics method is presented to study the sticking of H atoms in chemisorbed sites on the graphite surface in the collinear geometry. In the subsequent section, the different mechanisms for the formation of \hm\ are implemented into the Kinetic Monte Carlo (KMC) simulations. The sticking probability obtained by the dynamics method is included in the KMC code.  In the last section, we assess how this sticking changes the processes involved in the formation of H$_2$ \rm{and identify the predominant mechanisms to form \hm\ molecules as function of the gas and dust temperature.}\rm

\section{Sticking: Dynamical calculations}

In this section, the time dependent \rm{wave packets propagation method is}\rm\ presented to calculate the transmission probability for an H atom to overcome the chemisorption barrier. \rm{The time dependent dynamics method, that takes into account vibrational modes of the surface, is used to calculate the sticking probability of H atoms in chemisorbed sites on a graphite surface. Our  calculation are performed in the collinear approach. }\rm

\subsection{\label{sub:WPmethod}The wave packet propagation method}

This method solves for the time dependent Schrödinger equation:\\
\begin{equation}
i\hbar\frac{\partial\left|\psi(t)\right\rangle }{\partial t}=H\left|\psi(t)\right\rangle \label{eq:Schrodinger}\end{equation}
where H is the Hamiltonian of the H-graphite system which is the sum
of a kinetic operator and a potential operator and $\left|\psi(t)\right\rangle $
is the time-dependent wave function. This method consists of choosing
an initial wave function (at time t=0) to model the H atom in the
gas phase. The application of the evolution operator on the wave function
allows us to obtain the wave function at time $t+\Delta t$. The
time propagation is performed using a Lanczos method (\citealt{lanczos}).
\rm{The kinetic operator is applied on the wave function using a Fourier method (\citealt{kosloff1983})}\rm. The propagated wave function is analyzed by a flux analysis method
(\citealt{flux}) to obtain physical information such as the probability of transmission
of a potential barrier.

\subsection{The probability of transmission of the chemisorption barrier}

The interaction potential between the H atom and the graphite surface
is well known (\citealt{ferro2003}, \citealt{sha2002}, \citealt{jeloaica1999}).
\rm{This potential is represented on the figure~\ref{trans} as a function of the reaction coordinate z between the H atom and the graphite surface in the collinear approach. The main characteristics of the potential derived from the DFT code PWSCF included in the QUANTUM ESPRESSO package of (\citealt{PWSCF}), are the following:}\rm 
\begin{enumerate}
\item A surface reconstruction is observed: the H can be chemisorbed on
the graphite surface on top of a C atom with an energy bond of 0.76eV.
The C atom where H is chemisorbed, goes out of the surface plane. 
\item Chemisorption on the surface has an associated barrier of 0.2
eV. 
\end{enumerate}
The probability of transmission (Fig.\ref{trans}) of the chemisorption
barrier is calculated using the wavepackets propagation method (described
in section \ref{sub:WPmethod}) as a function of the incident energy
of the H atom coming in from the gas phase. The probability of transmission decreases with decreasing incident energy of the H atom. For energies
smaller than 0.2 eV (the value of the chemisorption barrier), the
H atom gets through the barrier by the tunneling effect.

\rm{The DFT calculations  are performed on a  2$\times$2 working cell. In this case, the chemisorption well is about 0.76eV and the chemisorption barrier is about 0.2eV. \cite{casolo2009} have shown that the chemisorption well is affected by the size of the working cell. For example, for a 5$\times$5 working cell, the chemisorption well is 12$\%$  higher than with a 2$\times$2 working cell and the chemisorption barrier is of 0.15 eV.  In our dynamics calculations, we consider a chemisorption barrier 0.2eV. The value of this barrier will affect the sticking probabilities and the transmission probability of the barrier. For a smaller barrier (\citealt{casolo2009}) the sticking probabilities and the transmission barrier calculated using dynamics method would be higher than the values calculated here in the energy domain between 0.15eV and 0.2eV. }\rm

\subsection{Sticking of H on graphite using a mixed classical-quantum
dynamics method}

\rm{The detail of the theory to study the sticking of H atom on a graphite surface using the mixed classical-quantum dynamics method is presented in the paper of \cite{morisset2010}. Here, we present a brief description.}\rm\ The hamiltonian used to describe the interaction of the atom with
the dynamical surface is:

\begin{equation}
H=T+V_{\mbox{tot}}+H_{\mbox{b}}\label{eq:Htot},
\end{equation}
 where $T$ is the kinetic operator, $V_{\mbox{tot}}$ is the gas-surface
potential operator, $H_{\mbox{b}}$ is the Hamiltonian for the lattice
vibration without coupling to an atom. In the mixed classical-quantum
method, it has the following form:

\begin{equation}
H_{\mbox{b}}=\sum_{i=1}^{N}\sum_{\mbox{j=1}}^{M}\left(\frac{1}{2}w_{\mbox{ij}}\left(Q_{\mbox{ij}}^{\mbox{2}}+P_{\mbox{ij}}^{\mbox{2}}\right)\right)\label{eq:hb},
\end{equation}
where $N$ and $M$ are the number of phonon modes and phonon bands,
respectively. $Q_{\mbox{ij}}^{\mbox{}}$ and $P_{\mbox{ij}}^{\mbox{}}$
are the positions and momenta of the vibrational modes of frequency
$w_{\mbox{ij}}$. In the previous papers of  Morisset \& Allouche (2008, 2010),
the Taylor series expansion of the potential $V_{tot}$ in terms of
the displacements of the lattice atoms allows one to write the potential
as the sum of the static potential $V_{0}(z)$ between H and the graphite
surface in the collinear approach plus an interaction potential between
the atom and the bath of phonons. The Taylor expansion is truncated
to the linear term in phonon coordinates, which is the one-phonon
exchange approximation. In the mixed classical-quantum approach, the
total gas-surface interaction can be written as:

\begin{equation}
V_{tot}=V_{0}(z)+\sum_{i=1}^{N}\sum_{\mbox{j=1}}^{M}g_{ij}(z)Q_{ij}\label{eq:vtot},\end{equation}
where $g_{ij}(z)$ \rm{represents the coupling terms between the motion of H and the bath of phonons with a vibrational mode of frequencies $\omega_{ij}$ and a polarization vector $\epsilon_{ij}$. To calculate from DFT calculations the $g_{ij}(z)$ terms, a model has been developed by \cite{morisset2008}. The $g_{ij}(z)$ terms are directly calculated of DFT phonon calculation in the harmonic approximation (\citealt{ashcroft1976}).  These calculations allow to obtain the dynamical matrixes D in the irreducible Brillouin Zone (IBZ). By diagonalization of D, the frequencies and polarization vector of the vibrational mode are obtained. The phonon dispersion of the H-graphite system is presented in the Figure 2 of the paper of \cite{morisset2008} along the high-symmetry directions in the IBZ. The phonon dispersion obtained by DFT phonon calculations are in good agreement with previous paper on the subject (\citealt{mohr2007})\rm

The dynamics method is performed through a mixed classical-quantum
approach. In this method, the H motion \rm(z coordinate between H and the graphite surface)\rm\ is treated quantum-mechanically
using the wavepackets propagation method (described in section \ref{sub:WPmethod}),
whereas each vibrational mode is treated classically. The time dependence
of the classical variables is obtained by solving the equations of
motion derived from the Hamilton equations:

\begin{eqnarray}
\dot{Q}_{ij} & = & \frac{\partial H_{eff}}{\partial P_{ij}},\nonumber \\
\dot{P}_{ij} & = & -\frac{\partial H_{eff}}{\partial Q_{ij}},\label{eq:hamilton}\end{eqnarray}
 where $H_{eff}$ is the effective Hamiltonian, given by the expectation
value of the Hamiltonian H :

\begin{equation}
H_{eff}=<\psi\mid H\mid\psi>.\label{eq:Heff}\end{equation}
 Initially, each vibrational mode obeys the equipartition theorem
(\citealt{kbt}): there is $\frac{k_{B}T}{2}$ of energy in each vibrational
mode where T is the surface temperature. The wave function is analyzed
using the reactive flux as described in \cite{morisset2010}, which allows one
to calculate the sticking probability as a function of the incident
energy of the H atom and the surface temperature T.

\subsection{Dynamics results}
\subsubsection{Transmission coefficients}
In previous studies, we calculated the probability for an atom to stick on a grain surface to be equal to the transmission coefficient to cross the barrier against chemisorption (\citealt{cazaux2002}, \citealt{cazaux2004}). We considered the barrier to be square, and obtained transmission coefficients that strongly depend on the energy of the incoming H atom (see  fig.\ref{trans}, right panel, dotted lines). The transmission coefficients are very small at energies below 0.2 eV ($\sim$2320K, which corresponds to the height of the barrier).  

However, the barrier against chemisorption is far from being square. Actually, DFT calculations show that this barrier varies from 0.79 \AA\ to 1.05 \AA\ in width, with a height of 0.2 eV (see  fig.\ref{trans}, left panel, \citealt{morisset2008}). The transmission coefficients obtained with such a barrier are shown as solid lines in fig.\ref{trans}, right panel. It is clear that the approximation of a square barrier underestimates the transmission coefficients by orders of magnitude. Already at energies of 1000~K, an H atom would have 10$^7$ times more chance to cross a realistic barrier as calculated by DFT than a square barrier. This effect has profound consequences for the sticking of H atoms at low gas temperatures, and will influence which processes govern the formation of H$_2$ and with what efficiency.

\subsubsection{Sticking}
In fig.\ref{stick}, we present the sticking of H atoms on graphite surfaces calculated with the mixed classical-quantum dynamics method in order to consider the dissipation of the H atom's energy through the different modes of the lattice. In the present calculation, N=11 phonon bands with M=61 vibrational modes have been included: the phonon bands of \rm{C-H (H chemisorbed on the graphite), the four phonon bands of C-C of higher frequencies, the longitudinal mode of C-C, and the three acoustic bands of C-C. Our dynamics calculation shows that the energy exchange between H and the vibrational modes of the substrate takes place by acoustic phonon modes of C-C.}\rm The chemisorption barrier (0.2~eV=2320~K)  governs the sticking mechanism: the sticking probability decreases with the incident energy for each surface temperature T$_{dust}$. At incident energies smaller than the chemisorption barrier, the sticking coefficient decreases with increasing T$_{dust}$. At low dust temperatures, the surface allows better dissipation of the energy of the incident atom, and therefore, facilitates the sticking. For example, the difference in sticking efficiency on surfaces of 10~K or on surfaces of 125~K can be a factor of 3 for incident energies around 1000~K. At incident energy higher than the chemisorption barrier, the variation of the sticking probability with the surface temperature is negligible.

\rm{Our calculation can be compared to experimental work by \cite{zecho2002}. For a surface temperature of 150K and with a beam centered about 0.2eV, these authors obtain a sticking probability of 0.5. In our dynamics calculation, in the same condition in temperature and energy, our sticking probability is about 0.6. Our results are in good agreement with the experimental work of \cite{zecho2002}.The other theoretical calculations on the subject  have been performed by  \cite{sha2005} and \cite{Kerwin2006} who obtain  a probability of 0.1 and 0.0025 in the same domains of energy and temperatures, respectively. These probabilities are smaller because these authors included in their dynamics calculations a coordinate to simulate the energy dissipation in the surface,  but do not explicitly take into account the vibrational modes of the substrate. Therefore, to study theoretically the sticking of an atom on a graphite surface, it is necessary to take into account the vibrational modes of the surface in the dynamics calculations.}\rm

\section{H$_2$ formation using Kinetic Monte Carlo simulations}
We use a  step by step Monte Carlo simulation  to follow the chemistry occurring on dust grains. \rm{This method allows us to follow each individual H atom arriving, binding and moving on the surface, and forming molecules through different mechanisms. However, this method does not give us informations on the energy of the atoms and formed molecules}.\rm\ The graphite surface is comprised of benzenic rings that allow the H atoms to physisorb or chemisorb, as presented in fig.~\ref{benzene}. H atoms can physisorb above each C atom, and on the bridge between 2 C atoms (position E). However, recent studies showed that H atoms cannot physisorb at the center of the ring (\citealt{ferro2002}). The physisorbed energy has been chosen to be 43 meV  from \cite{ghio1980} and the barrier between 2 physisorbed sites is almost negligible (5-7 meV, \citealt{bonfanti2007}).  H atoms can chemisorb on top of C atoms with an associated barrier of 0.2~eV and an energy of 0.76 eV. If an atom is already present on a ring, a second atom, with a spin opposite to the adsorbed atom, can become chemisorbed in a para configuration \rm{without a barrier (\citealt{rougeau2006}), making a para-dimer. H atoms can also become chemisorbed in ortho and meta configurations, but these processes are associated with a barrier of 0.26 and 0.16~eV respectively (Rougeau et al. 2006)}\rm\. The para-dimer is the favorable configuration to form \hm\ because it is barrier-less compared to the other configuration. A third atom arriving on this dimer can create an H$_2$ molecule without a barrier (\citealt{bachellerie2007}). If an atom from the gas phase arrives on an adsorbed atom (a monomer), then there is a barrier of 9 meV, as derived by \cite{morisset2004}. 

Species that are accreted from the gas phase arrive at a random time and location on the dust surface. This arrival time depends on the rate at which gas species collide with the grain. This rate of accretion can be written as:
\begin{equation}
R_{acc} = n_H v_H  \sigma  S, 
\end{equation}
where n$_H$ and v$_H$ are the density and velocity of the H atoms in the gas, $\sigma$ is the cross section of the dust particle and $S$ is the sticking coefficient of the H atoms with the dust. The sticking in chemisorbed sites has been discussed in the previous section. This coefficient strongly depends on the form of the barrier to chemisorb, but is also somewhat sensitive to the surface temperature since energy transfer to or from the phonon can be important to loose excess energy. In figure \ref{stick}, left panel, different sticking probabilities for the physisorption case are presented as a function of gas temperature (T$_{gas}$). In this figure, S1 represents the sticking probability derived by \cite{cuppen2010} based on a soft cube model:
\begin{equation}
S_{phys}=(1+a_1\sqrt{T_{gas}+T_{grain}}+a_2 T_{gas}-a_3 T_{gas}^2)^{-1},
\end{equation}
with a$_1=4.2 \times 10^{-2}$ K$^{-1/2}$, a$_2 =2.3  \times 10^{-3}$ K$^{-1}$, and a$_3=1.3 \times 10^{-7}$ K$^{-2}$.
S1 decreases with the energy of the gas. At 100~K, S1=0.8 while it becomes 0.2 at 1000~K. At low T$_{gas}$, S1 is sensitive to the temperature of the dust. In this model, the sticking probability of H in physisorbed sites on graphite is important.  S2 represents the sticking probability obtained by \cite{buch1989}. This model is based on the quantum mechanical perturbation theory.  This calculation includes a microscopic description of the solid structure and vibrations. S2 is very weak, and varies between 0.02 and 10$^{-4}$ on the energy domain T$_{gas}$=50-550~K.  

Atoms arriving from the gas phase on top of a carbon atom can either be physisorbed (with a sticking efficiency S$_{phys}$) or be chemisorbed (with a sticking efficiency S$_{chem}$).  In our model, because of  the high barrier to access chemisorbed sites, H atoms mostly arrive from the gas phase in physisorbed sites for low temperature gas.  
 
The species that are present on the surface can go back into the gas phase though evaporation. The evaporation rate of H atoms in physisorbed or chemisorbed sites can be written as:
 \begin{eqnarray}
R_{evap(H_p)}& =& \nu_p \times \exp{(-\frac{E_p}{k_BT})},\\ \nonumber 
R_{evap(H_c)} &=& \nu_c \times \exp{(-\frac{E_c}{k_BT})}, 
\end{eqnarray}
where E$_p$ and E$_c$ are the binding energies of the H atom in a physisorbed or chemisorbed site, respectively, and $\nu_p$ and $\nu_p$ are the oscillation factors of the atoms in the physisorbed and chemisorbed sites taken as  $\nu_p$=10$^{12}$s$^{-1}$ and  $\nu_c$=10$^{13}$s$^{-1}$ .
 
The species that arrive at a location on the surface can move randomly through tunneling effects and thermal hopping. The diffusion rates for an atom to go from a physisorbed to another physisorbed site (R$_{pp}$), and for an atom to go from a physisorbed site to a chemisorbed site (R$_{pc}$)  can be written as:
\begin{eqnarray}
R_{pp} &=& \nu_p \times \exp{(-\frac{E_{pp}}{k_B T})},\nonumber \\
R_{pc} &=& \nu_p \times  P_{pc},
\end{eqnarray}
where P$_{pc}$ is the probability for the atom to go from a physisorbed site  to a chemisorbed site by tunneling effects or thermal hopping, as described in \cite{cazaux2004}. When thermal hopping dominates, which is usually the case for high temperatures, or when the barrier is low, this probability can be written as $P_{pc}=\exp({-\frac{E_{pc}}{k_B T}})$,  where $E_{pc}$ is the energy of the barrier between the physisorbed and the chemisorbed site. This barrier is 0.2~eV, which means that physisorbed H atoms cannot directly reach chemisorbed sites. However, if a chemisorbed atom is already present, then the barrier in its associated para-site disappears. In this case, the physisorbed H atom can reach these para-sites without a barrier and consequently will hop thermally in chemisorption.

The mechanisms included in our KMC simulations are the mechanisms M1 to M5 described in
the introduction. \cite{dumont2008} have included in their KMC calculations formation of  ortho, para, meta, and 5 and S dimers (see fig~\ref{benzene}, right panel). The H atoms can diffuse from one dimer position to another. These simulations explained the occurrence of two peaks in the thermal recombinative desorption of molecular H$_2$  from clean graphite surfaces, after chemisorption of H  atoms. The first peak (450~K) is essentially due to the para-dimer desorption whereas the second peak (560~K) is due to the ortho-dimer desorption. The others mechanisms included in their simulation contribute to the \hm\ desorption but with a small amount. In this work we concentrate on \hm\ formation on dust grains with temperatures smaller than 150~K. The mechanisms considered by \cite{dumont2008} are not relevant for the range of temperatures considered in this study since diffusion of chemisorbed H atoms as well as thermal recombinative desorption occur at much higher temperatures. In this study we use similar mechanisms as the ones considered in \cite{cuppen2008}, and also include the formation of \hm\ with 2 physisorbed atoms.

The mechanisms to form \hm\ are associated with very high barriers, if they involve a chemisorbed H atom in a monomer (mechanisms 2 and 4), and are barrier-less if chemisorbed H atoms in para-dimers are involved. We perform several simulations for the formation of H$_2$ on graphitic surfaces with various grain and gas temperatures. We first consider the barrier to enter chemisorbed sites as being square, and then consider more realistic barriers obtained by DFT calculations. Finally, we also take into account the effect of the phonons on the sticking of atoms in chemisorbed sites.

In our simulations, we calculate the total sticking coefficient times the total \hm\ efficiency, S$\times\epsilon$. This product can be written:
\begin{equation}
S\times\epsilon= S_{phys}\times\epsilon_{phys}+S_{chem}\times\epsilon_{chem}, 
\end{equation}
 where S$_{phys}\times\epsilon_{phys}$ is the product of the sticking coefficient in physisorbed sites times the \hm\ efficiency involving physisorbed atoms, and  S$_{chem}\times\epsilon_{chem}$ is the product of the sticking coefficient in chemisorbed sites times the \hm\ efficiency involving chemisorbed atoms. The efficiency of \hm\ formation is written as:
 \begin{equation}
\epsilon= \frac{2\times n_{\rm{H_2}}}{n_{\rm{H}}},
\end{equation}
where n$_{H_2}$ is the number of \hm\ molecules formed, and n$_H$ the number of H atoms that arrived on the surface.

\subsection{Effect of the barrier on H$_2$ formation}
The transmission probability for H atoms to overcome the chemisorption barrier (0.2 eV) \rm{in the collinear approach}\rm\ is calculated following two different approximations:\\
	1) the chemisorption barrier is considered as a square barrier (figure~\ref{trans}, right panel, dashed lines). The transmission probability is presented in figure~\ref{trans}, left panel (dashed lines),\\
	2) the chemisorption barrier is obtained by DFT calculations (figure~\ref{trans}, right panel, solid lines) using the wavepacket propagation method described in paragraphs 2.1 and 2.2, and the transmission probability is presented in figure~\ref{trans}, left panel  (solid lines).

The simulations for the formation of H$_2$ considering a square barrier and a barrier obtained by DFT calculations to enter chemisorbed sites are shown in fig.\ref{h2bar}. \rm{The square barrier model has been used in previous studies (\citealt{cazaux2002}, \citeyear{cazaux2004}). The objective of this study is to understand the influence of the  form of the chemisorption barrier on the \hm\ formation using the KMC simulations.}\rm\ For the \hm\ formation, mechanisms M1 to M5 are included in the KMC simulations. In this figure~\ref{h2bar}, $\epsilon_1$ represents the efficiency of H$_2$ formation when a square barrier is considered, while $\epsilon_2$ represents the same efficiency when the barrier is obtained by DFT calculations. We considered temperatures of the gas of 100, 500, 1000 and 2000~K with grain temperatures from 8 to 90 K. Our results show that:
At low dust temperatures (T$_{dust}$<15~K for barrier 1 and < 20~K for barrier 2), the formation of \hm\ involves physisorbed atoms, either with the mechanism M1 (2 physisorbed atoms) or with the mechanism M5 (one physisorbed atom entering a chemisorbed site occupied by an H atom in a dimer). The efficiency to form \hm\ at these temperatures is on the order of 100$\%$, and therefore the product S$\times \epsilon$ is equal to S$_{phys}$. At dust temperatures higher than $\sim$15~K, the physisorbed atoms evaporate, and molecular hydrogen is formed with chemisorbed atoms. The type of barrier considered has a very important effect on the formation of H$_2$ since it determines the sticking of H atoms in chemisorbed sites. At low gas temperatures, the H atoms have a low probability to enter the chemisorbed sites, and therefore the formation efficiency is low. \hm\ is formed through the direct Eley-Rideal mechanism when an H atom from the gas arrives on a dimer on the surface (M3). Once a first H atom sticks to one chemisorbed site, the second one sticks to a para-site without a barrier, and a third H atom coming on the dimer can form an \hm\ molecule without a barrier. The atom that stays on the surface is used to make another dimer, again without a barrier, and again another H$_2$ molecule can be formed. With this process,  $S\times\epsilon$ is orders of magnitude higher than the sticking alone. As the gas temperature increases, the barrier to become chemisorbed becomes easier to overcome, and \hm\ can form through the direct Eley-Rideal mechanism when an H atom from the gas arrives on a single H on the surface (M2).

\subsection{Effects of the surface dynamics on the formation of  \hm}
We used Kinetic Monte Carlo simulations including \rm{mechanisms M1 to M5}\rm\ to calculate the formation of \hm\ on dust surfaces of 10 and 125~K as function of the gas temperature. The atoms present in the gas phase arrive on the surface with an energy that allows them either to stick in chemisorbed sites (if the barrier to chemisorption can be overcome), or to stick in physisorbed sites (if atoms can thermalize in physisorbed sites), or to bounce back to the gas phase. For the two surface temperatures considered, the sticking in chemisorption has been calculated previously (section 2.4.2), and takes into account the effect of surface dynamics. However, the sticking in physisorption varies strongly from one study to another. In this work we consider strong sticking in physisorption (\citealt{cuppen2010}), and weak sticking in physisorption (\citealt{buch1989}), and calculate for each case the product of sticking times efficiency for \hm\ formation, $S\times\epsilon$. 

\subsubsection{Strong sticking in physisorption.}
In fig.\ref{phon1} (left panel), we show the product $S\times\epsilon$ for surface temperatures of 10 and 125~K, and for gas temperatures varying from 100 to 3000~K. This product is the sum of the \hm\ formation efficiency times sticking involving  physisorbed atoms, S$_{phys}\times\epsilon_{phys}$, and involving chemisorbed atoms, S$_{chem}\times\epsilon_{chem}$. In these calculations, we consider a sticking coefficient for physisorption, S$_{phys}$, such as the one in \cite{cuppen2010}, and shown in fig~\ref{stick}. For  T$_{dust}$=10~K,  S$\times\epsilon$ is very large at low gas temperatures and decreases as gas temperature increases. Because of the high sticking probability to physisorb, the formation of \hm\ involves physisorbed atoms and therefore $S\times\epsilon$ depends on S$_{phys}$. At gas temperatures lower than 500~K, \hm\ is formed through the association of two physisorbed atoms (mechanism M1), while for higher gas temperatures, physisorbed atoms enter chemisorbed sites and associate with H atoms in a dimer (mechanism M5). For T$_{dust}$=125~K, S$\times\epsilon$ is very small at low gas temperatures and increases as the gas temperature increases. Because of the high surface temperature, physisorbed atoms do not stay on the surface and the formation of \hm\ is insured by chemisorbed atoms.  The formation of \hm\ proceeds through direct Eley-Rideal mechanisms involving a chemisorbed H atom in a dimer (M3), for T$_{gas}$<1000~K, and involving an chemisorbed H atom in a monomer for higher T$_{gas}$ (M2). For the two different surface temperatures considered, the mechanisms to form \hm\ are different since they either involve physisorbed atoms (for T$_{dust}$=10~K)  or chemisorbed atoms (for T$_{dust}$=125~K). However, the effects of surface dynamics on the formation of \hm\ can be appreciated only by comparing the formation of \hm\ involving chemisorbed atoms. For this purpose we represent  in fig.\ref{phon1} (right panel) the different contributions for the formation of \hm\ that involve only chemisorbed atoms; S$_{chem}\times\epsilon_{chem}$. The different mechanisms for the formation of \hm\ that involve only chemisorbed atoms are shown as dashed lines for T$_{dust}$=125~K and solid lines for T$_{dust}$=10~K. For T$_{gas}$<1000~K, \hm\ forms through Eley-Rideal mechanisms involving a chemisorbed H atom in a dimer (M3), while for higher T$_{gas}$, H atoms can access chemisorbed sites easily even  with a barrier, and \hm\ is insured through Eley-Rideal mechanisms involving a chemisorbed H atom in a monomer (M2). The product  S$_{chem}\times\epsilon_{chem}$ slightly depends on the surface temperature, but the effect of the surface dynamics cannot be assessed. Indeed, the mechanism involving chemisorption depends also on the physisorption since many physisorbed atoms can travel on the surface and become chemisorbed. Therefore, as long as some H atoms can become physisorbed and subsequently reach the chemisorbed sites, it is not possible to really isolate the impact of surface dynamics on the formation of \hm.

\subsubsection{Weak sticking in physisorption}
In fig.\ref{phon2}, we present the same calculations as previously, but we now consider a very small sticking coefficient for physisorption, S$_{phys}$, as described in \cite{buch1989}. In this case, S$_{phys}$=0.02 at low gas temperatures and S$_{phys}$ decreases exponentially as T$_{gas}$ increases (see fig.\ref{stick}). Our results show that because atoms do not stick in physisorbed sites, \hm\ formation is insured by mechanisms involving only chemisorbed atoms. For T$_{gas}$<1000~K, \hm\ forms through Eley-Rideal mechanisms involving a chemisorbed H atom in a dimer (M3), while for higher T$_{gas}$, gas phase H atoms can enter chemisorbed sites even in the presence of a barrier, and \hm\ formation is insured through Eley-Rideal mechanisms involving a chemisorbed H atom in a monomer (M2). The total product $S\times\epsilon$ is equal to S$_{chem}\times\epsilon_{chem}$, and the effects of the surface dynamics on the sticking in chemisorption can be addressed. If the surface is flat (represented as small dashed lines), then S$_{chem}$ is smaller (see fig.\ref{stick}) and the product is therefore smaller.  When the surface dynamics are considered, the sticking  $S_{chem}$ is more efficient, and therefore the product  $S\times\epsilon$ is slightly higher. We also note that the effect of the surface temperature on the product  $S\times\epsilon$ is almost negligible, as shown by the curves for T$_{dust}$=10~K (solid) and T$_{dust}$=125~K (small dashed).

\section{Discussion and conclusions}
In this work, we have studied the influence of the sticking probability on the formation of  H$_2$ \rm{using Kinetic Monte Carlo simulations. In our model, mechanisms M1 to M5 (described in the introduction) for the \hm\ formation are included. We have determined the predominant mechanisms to form \hm\ as function of gas and dust temperatures.

First, we assumed the sticking probability of H atoms to be equal to the transmission coefficient to cross the barrier against chemisorption. To address the importance of the shape of the barrier, we consider a square barrier, and a barrier obtained by DFT calculations. For the latter, wavepacket propagation calculations along the coordinate between the H atom and the graphite surface (collinear approach) are performed to determine the transmission probabilities of the chemisorption barrier. Our results show that if a square barrier is considered, the transmission probabilities are underestimated by many orders of magnitude at low gas temperatures (T$_{gas}$$\le$2000K). The impact of the form of the barrier is visible on the results of the KMC calculations simulating the formation of \hm. Our results show that the shape of the chemisorption barrier (square or obtained by DFT) has a strong impact on the \hm\ formation for T$_{dust}$ $\ge$ 15~K. This impact increases as gas temperatures decrease. In conclusion, the formation of \hm\ calculated with barriers obtained by DFT is much more efficient than with square barriers. In these temperatures domain, \hm\ formation involves chemisorbed atoms. For T$_{dust}$$\le$15~K, \hm\ is formed by a physisorption mechanism. In this case, the form of the chemisorption barrier is not important.

Second, we investigated the sticking of H atoms in chemisorbed sites taking into account the vibrational modes of the surface. For this we used a mixed classical-quantum dynamics method where the motion of the H atom is treated quantum-mechanically, while the vibrational modes of the surface are treated classically (\citealt{morisset2008}). The sticking probability which take into account the vibrational modes of  the surface can be compared to the transmission probability of the barrier. The sticking probability is higher by a factor of 5 for T$_{gas}$ $\sim$1500K, but is lower by a factor 2 (for T$_{dust}$=10~K) to 5 (for T$_{dust}$=125~K) around T$_{gas}$ $\sim$1000K. This calculation show that the energy exchange between the atom and the bath of phonon is higher than transmission probability of the chemisorption barrier. These sticking probabilities are calculated as function of T$_{gas}$, for T$_{dust}$=10~K and T$_{dust}$=125~K. We implement these sticking coefficients S$_{chem}$ in our KMC  simulations, and show that for  T$_{dust}$=125~K, \hm\  is formed through a chemisorption mechanism ( involving dimers for T$_{gas}$$\le$ 1000~K and monomers for  T$_{gas}$$\ge$ 1000~K ). For T$_{dust}$=10~K, on the other hand, \hm\ formation involves physisorbed atoms if the sticking in physisorption is high. In this case the efficiency to form \hm\ depends of the sticking probabilities of H atom in a physisorption site.  \rm

The sticking in physisorption influences the formation of H$_2$ at low dust temperatures ($\le$15~K). In this study, we considered 2 possible sticking coefficients in physisorption S$_{phys}$. 1) The sticking  derived in \cite{cuppen2010}, which is valid for metals and is close to unity for low gas and dust temperatures. 2) The sticking derived by \cite{buch1989},  which shows that the probability of sticking for H atoms on graphite is on the order of a few percent (\citealt{lepetit2011}, \citealt{medina2008}). The influence of this S$_{phys}$ on the product $S\times \epsilon$ is very important at low dust temperatures. If the sticking is taken as in 1), then \hm\ is formed through LH mechanisms with physisorbed atoms, and $S\times \epsilon$ is very high $\sim$ S$_{phys}$. On the other hand, if the sticking in physisorption is chosen to be as in 2), the formation of \hm\ involving physisorbed atoms is not efficient, and the formation of \hm\ is insured by chemisorbed H atoms. In this case, the product $S\times \epsilon$ depends on S$_{chem} \times \epsilon_{chem}$. \rm{Our KMC simulations show that  once the vibrational modes of the H-graphite system are taken into account in the sticking, the efficiency to form \hm\ is increased by a factor 2. This behavior is not surprising since the sticking probability obtained by dynamics calculations is higher than the value of the transmission probability of the chemisorption barrier. }\rm

\rm{In the ISM, the \hm\ formed are ro-vibrationally excited.  The excitation of the molecule is due to the energy transfer of the kinetic energy towards the substrate, the translational energy and the excitation of the molecule. The ro-vibrational excitation of the molecule allows the determination of the mechanisms to form \hm\ (\citealt{lemaire2010}, \citealt{islam2007}). Different mechanisms have been studied theoretically by several groups (\citealt{rutigliano2001}, \citealt{ree2002}, \citealt{sha2002},\citealt{morisset2003}, \citealt{morisset2004}, \citealt{morisset2005}, \citealt{martinazzo2006phys},\citealt{bachellerie2007}). In our KMC simulations it is not possible to extract the energy of the \hm\ formed. These simulations only allow us to follow the number of molecules formed through different mechanisms (M1 to M5 described in the introduction). In this study, we have determined the predominant mechanism to form \hm\ as a function of the dust and gas temperatures.}\rm

Recently, \cite{cuppen2010} performed similar work on the formation of \hm\ and the influence of the sticking on the product  $S\times \epsilon$. These authors concentrated on the sticking in physisorbed sites (and we used their results in this work), but assumed that the atoms from the gas phase had to thermally overcome the barrier against chemisorption to become chemisorbed. \rm{In this sense, the H atoms coming from the gas phase do not stick in chemisorbed sites at low T$_{gas}$ (S$_{chem}$= 0.03, 0.0001 and 10$^{-8}$ for T$_{gas}$= 500, 200 and 100~K respectively). Therefore, our results are similar to \cite{cuppen2010}  for cold dust, when physisorbed atoms are involved to form \hm. However, on warm dust grains ($\sim$ 100K), chemisorbed atoms are needed to form \hm\ and the sticking of H atoms in chemisorbed sites sets the formation of \hm. In this case, our results diverge from \cite{cuppen2010} for gas temperatures lower than 500~K, since  we consider that H atoms from the gas phase can tunnel through the important barrier against chemisorption}. This process makes the sticking of H atoms in chemisorbed sites still efficient at low T$_{gas}$. Once these chemisorbed sites are populated, \hm\ can form efficiently through barrier-less routes involving the creation of dimers. Therefore, our results show that the formation of \hm\ is efficient also for intermediate gas and dust temperatures (T$_{dust}$>15K and 1000~K>T$_{gas}$>100~K).\rm 

Our results suggest that the formation of H$_2$ remains efficient in regions where gas and dust are warm. \rm{This is true if the time for H atoms to enter chemisorbed sites is shorter than other routes to form \hm\ (i.e. dense and warm medium). Therefore, the grain surface route proposed in this study should be compared to other routes to form \hm. }\rm\ Environments where grain surface chemistry would dominate to form \hm\ include photon dominated regions (PDRs, \citealt{hollenbach1999}), X-ray dominated regions (XDRs, \citealt{meijerink2005}), hot cores (\citealt{caselli1993}), and the early universe (\citealt{cazaux2004b}, \citeyear{cazaux2009}). Particularly PDRs and XDRs enjoy regions of upto $10^{22}$ and $10^{24}$ cm$^{-2}$, respectively, where rapid photo-dissociation and chemical removal of H$_2$ requires it to be formed efficiently on dust grains in order to drive ion-molecule chemistry and H$_2$ line emission (\citealt{habart2004}). In PDRs and XDRs gas temperatures can be as high as $10^3$ K (\citealt{meijerink2007}), values for which the traditional sticking coefficient of \cite{hollenbach1979} decreases rapidly. At early cosmic epochs, for redshifts larger than 10, the first grain surfaces are expected to be warm, at least 30 K, due to the strong cosmic microwave background. Furthermore, the low abundances of metals like carbon and oxygen cause gas temperatures to be $10^2-10^{3.5}$ K and to be set mostly by H$_2$ and HD cooling (\citealt{glover2008}). Our results indicate that chemisorption effects allow dust grains to act as catalysts for H$_2$ formation under such hostile primordial conditions.

\subsection*{ACKNOWLEDGMENTS}
The authors which to thank the anonymous referee for her/his constructive comments. S. C. is supported by the Netherlands Organization for Scientific Research (NWO). The authors wish to thank CINES for allowing them to use its computational facilities. This work is supported by the Euratom-CEA Association, in the framework of the F\'ed\'eration de Recherche Fusion par Confinement Magn\'etique,  and by the Agence Nationale de la Recherche (ANR CAMITER ANR-06-BLAN-0008-01).

\clearpage

\begin{figure}
\includegraphics[width=0.5\textwidth]{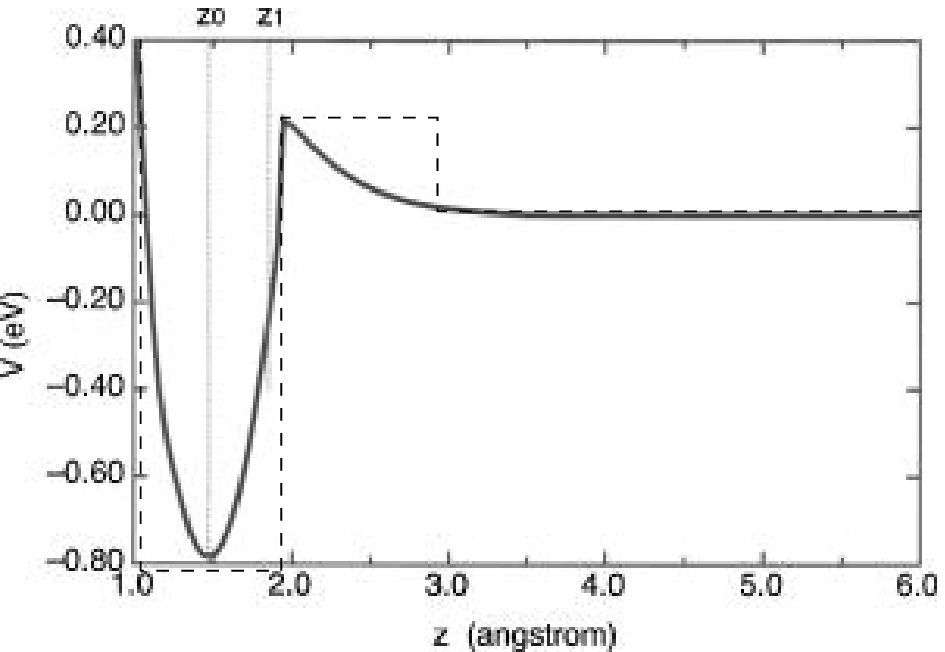}
\includegraphics[width=0.5\textwidth]{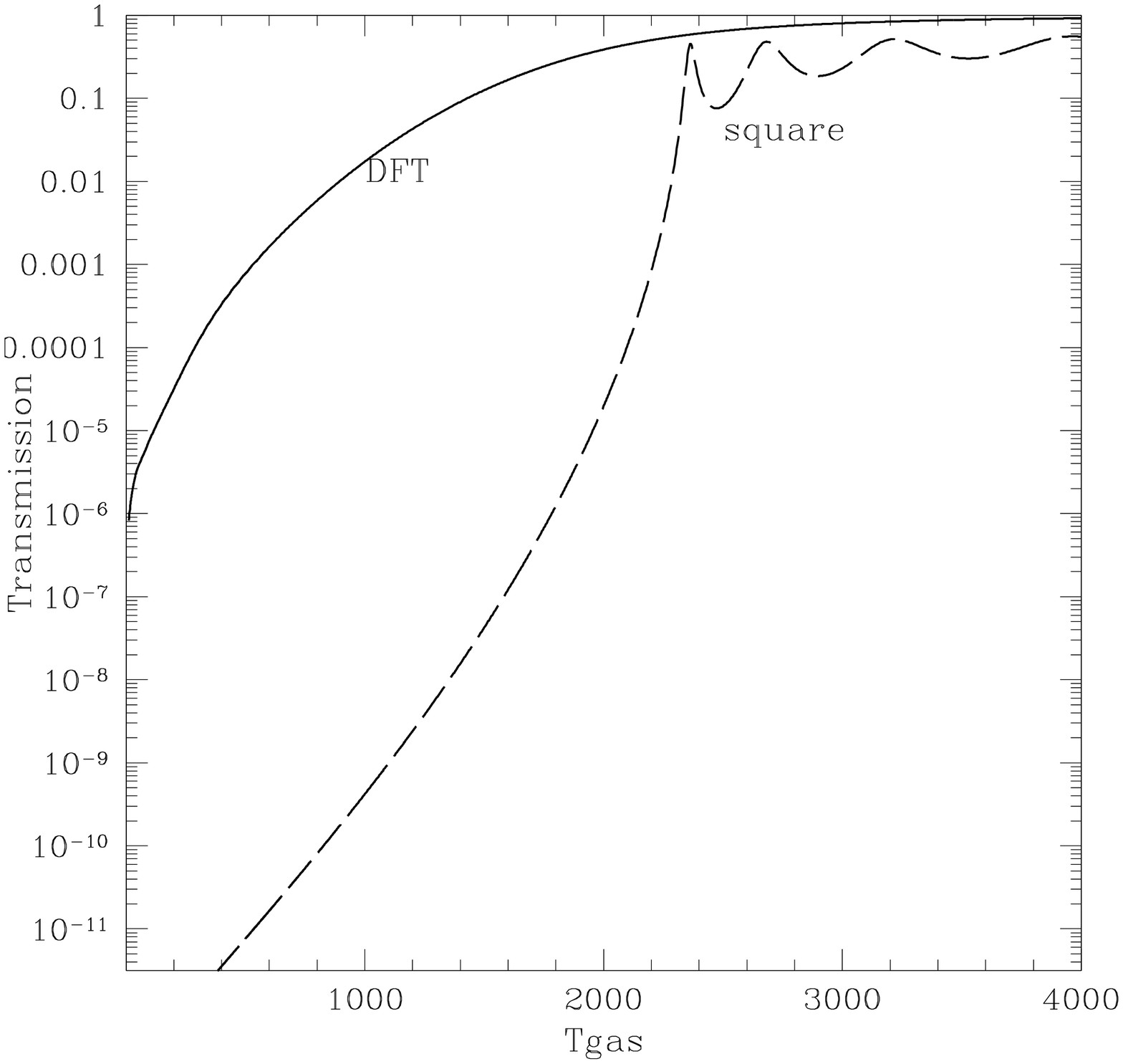}
\caption{Transmission coefficients for a square barrier approximation (dotted lines) and for a barrier obtained by DFT calculations (solid lines). Left panel: the different barriers considered. Right panel: transmission coefficients for H and D atoms versus gas temperature.}
\label{trans}
\end{figure}

\begin{figure}
\includegraphics[width=0.5\textwidth]{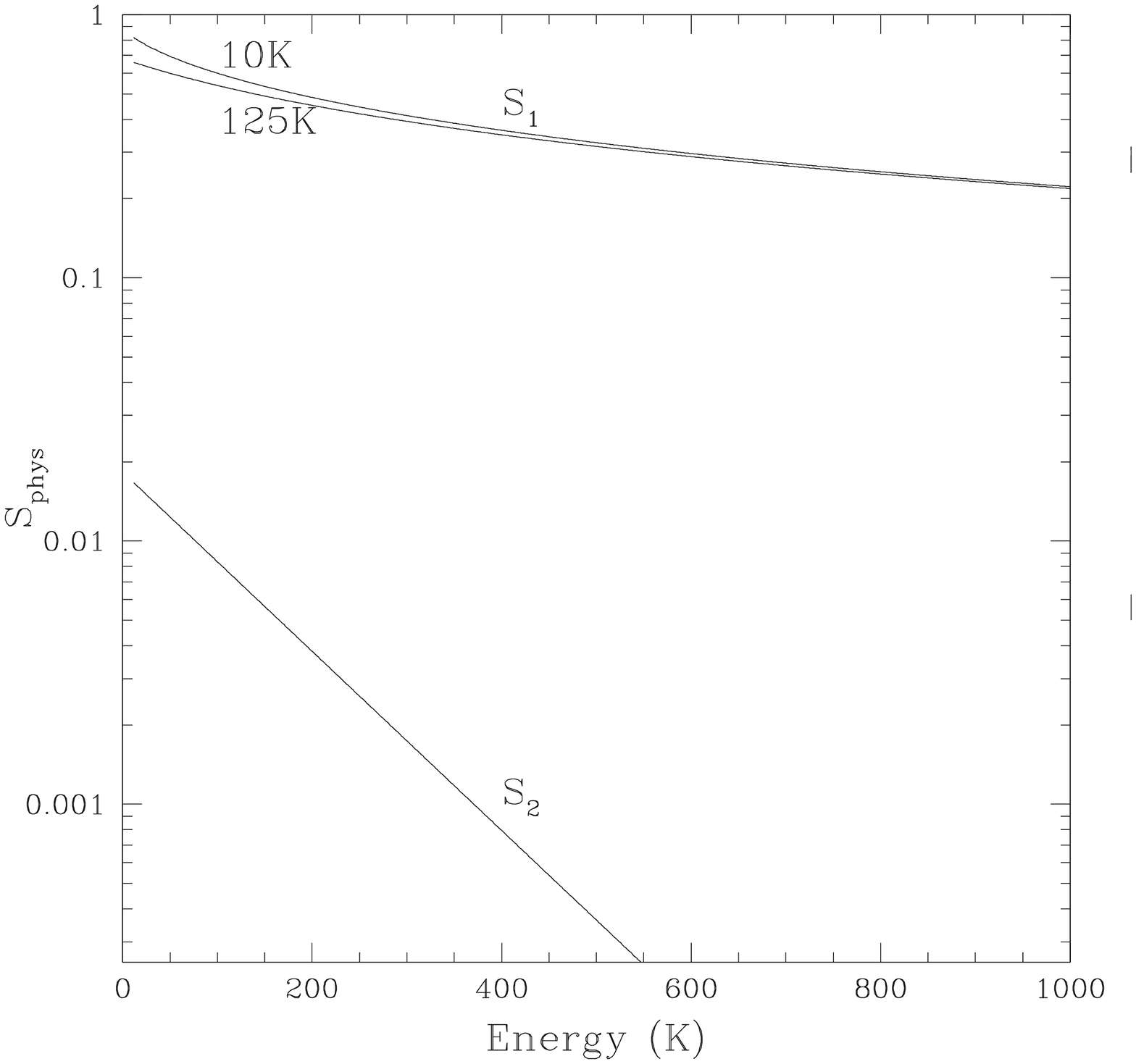}
\includegraphics[width=0.5\textwidth]{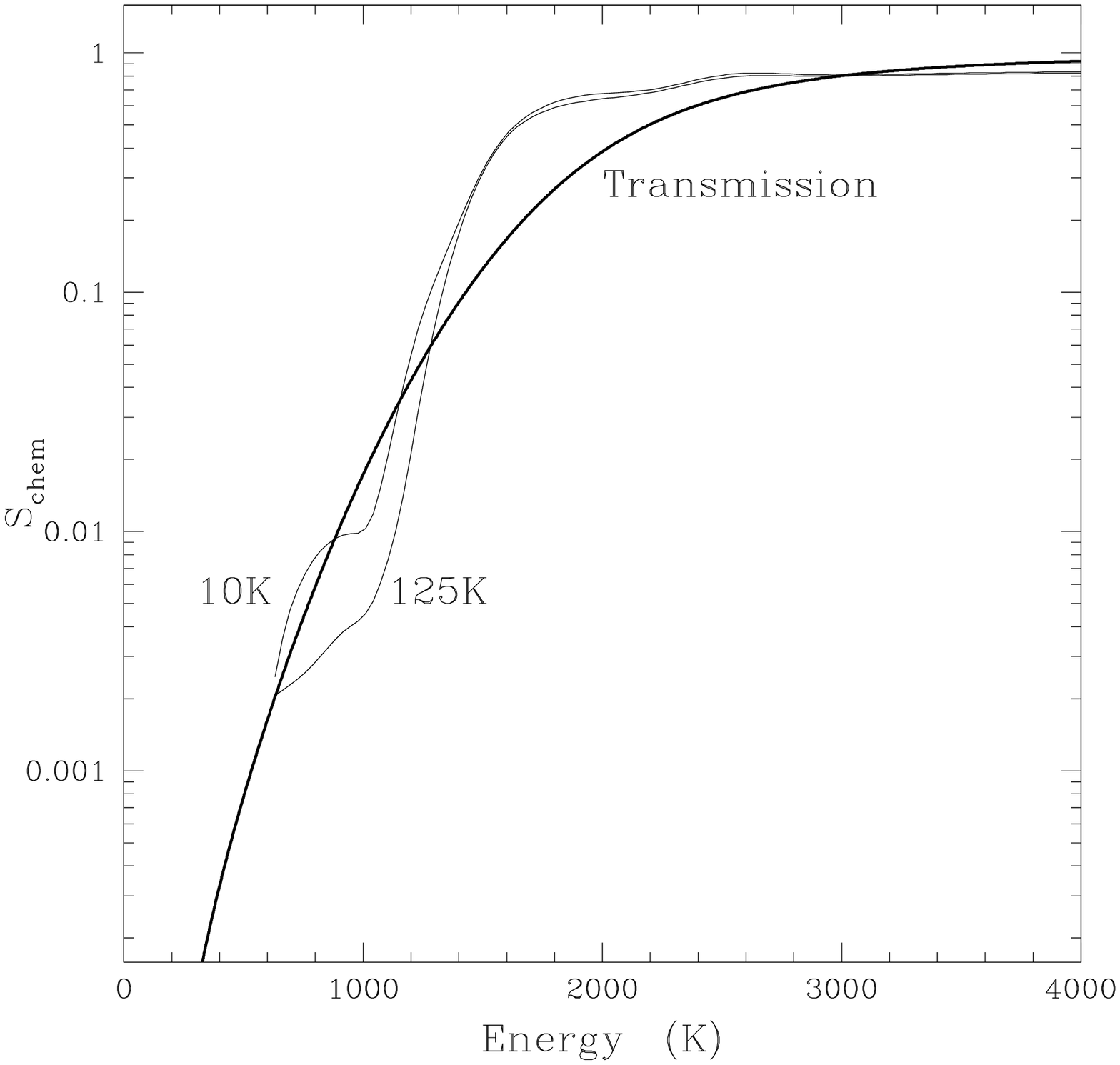}
\caption{Left panel: sticking probability in physisorbed sites as function of the energy of the incoming H atom according to \cite{cuppen2010} (S1, for T$_{dust}$=10 and 125~K) and \cite{buch1989} (S2). Right panel: trapping probabilities (transmission) and sticking in chemisorbed sites as function of the energy of the incoming H atom. The transmission coefficient to overcome a barrier derived from DFT calculations is shown as well as the sticking coefficient when the phonons are considered, for T$_{dust}$=10 and 125~K.}
\label{stick}
\end{figure}

\begin{figure}
\includegraphics[width=0.5\textwidth]{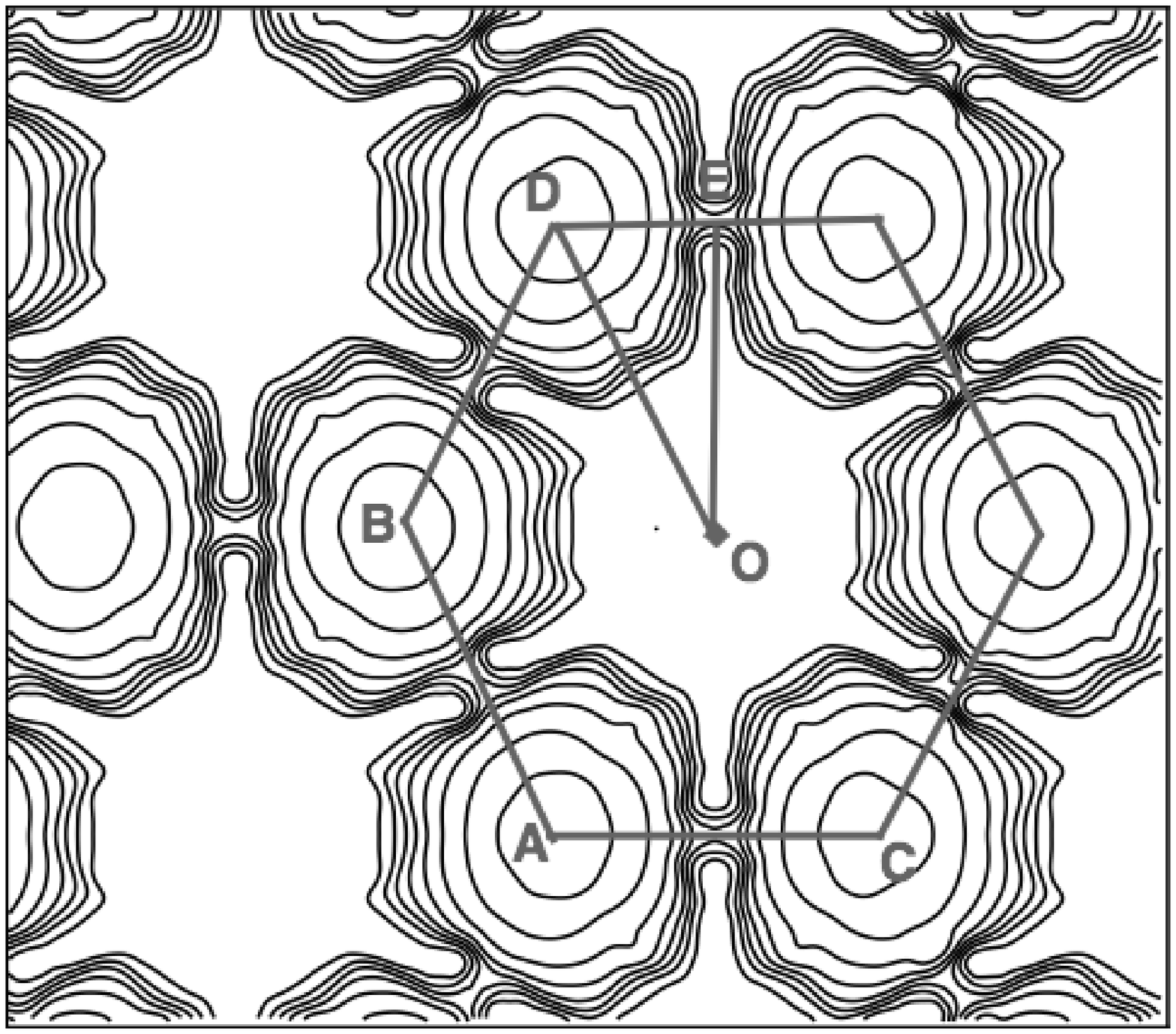}
\includegraphics[width=0.5\textwidth]{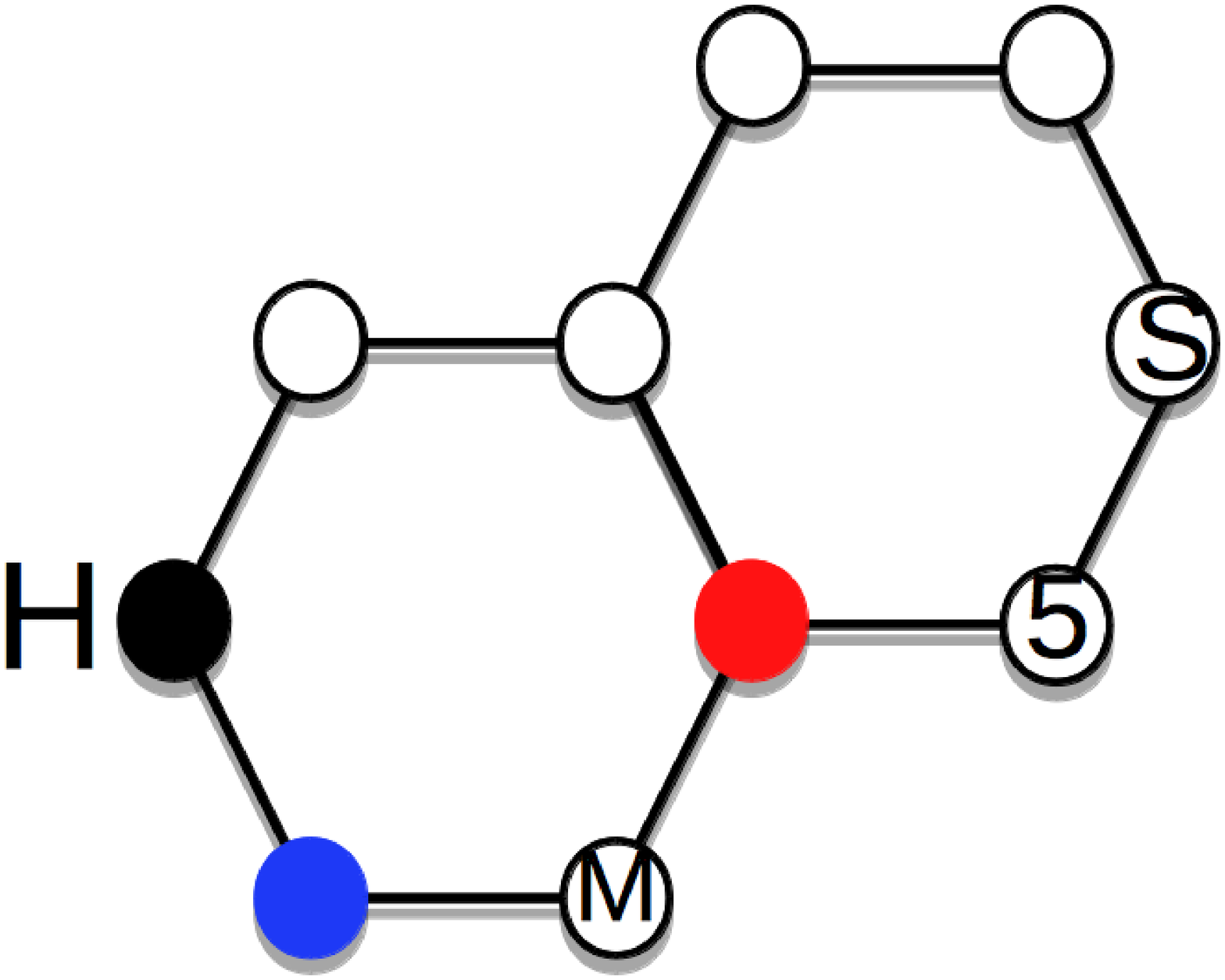}
\caption{Left panel: Contour map of the potential energy surface as function of the H position on the graphite. The energy difference between two consecutive contour lines is 0.3 eV. A, B, C and D are the positions of a barrier between two chemisorption sites. O is the center of the hexagon. H atoms can chemisorb  and physisorb on top of carbon atoms (A, B, C and D), while they can physisorb also in other locations such as a bridge (E). Right panel:  stable configuration for the second H atom: ortho (closer H atom on the hexagon), meta (M) and para (opposite H atom on the hexagon)  configurations and 5 and S (second cycle). (\citealt{dumont2008}) }
\label{benzene}
\end{figure}

\begin{figure}
\includegraphics[width=0.5\textwidth]{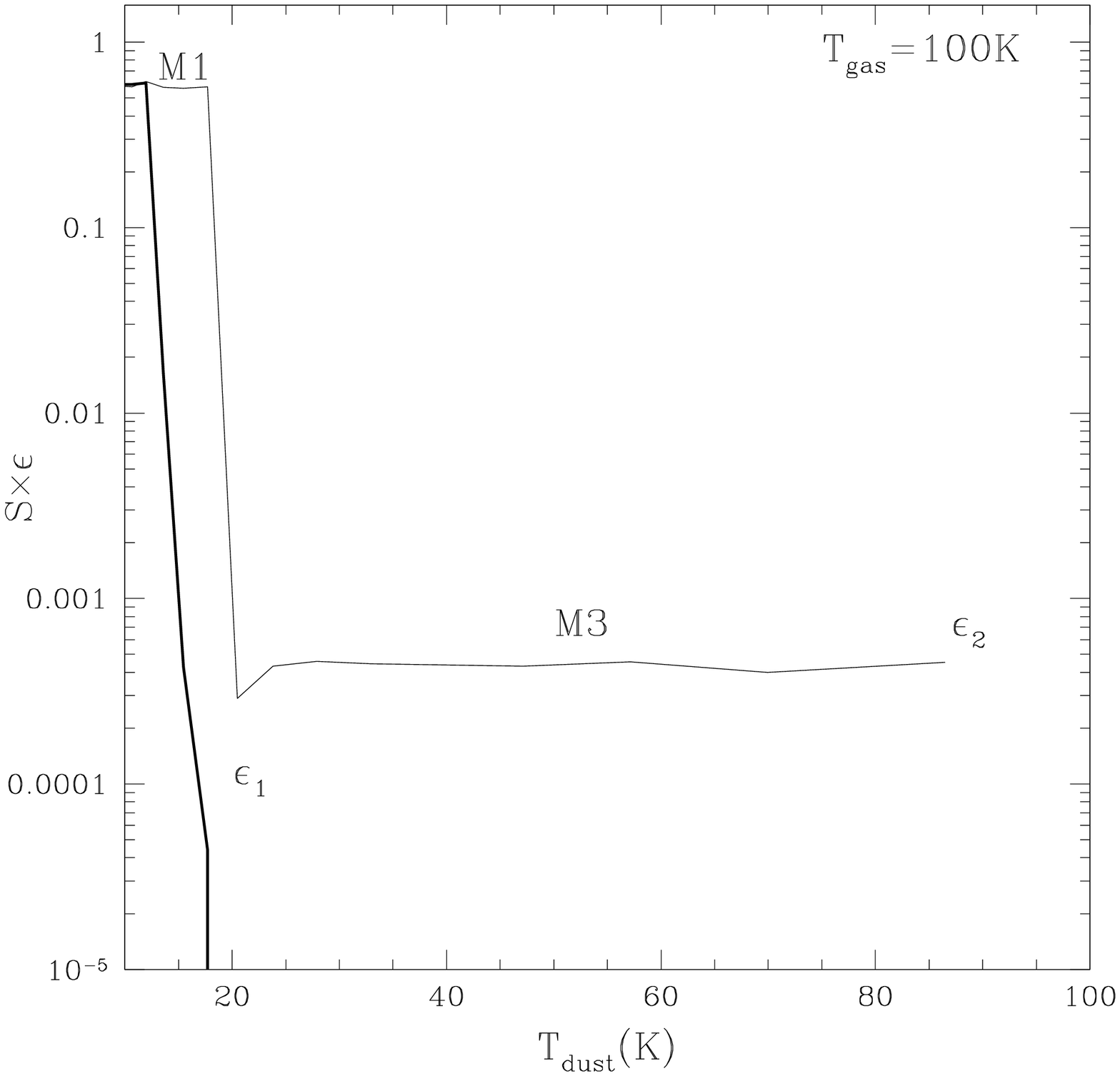}
\includegraphics[width=0.5\textwidth]{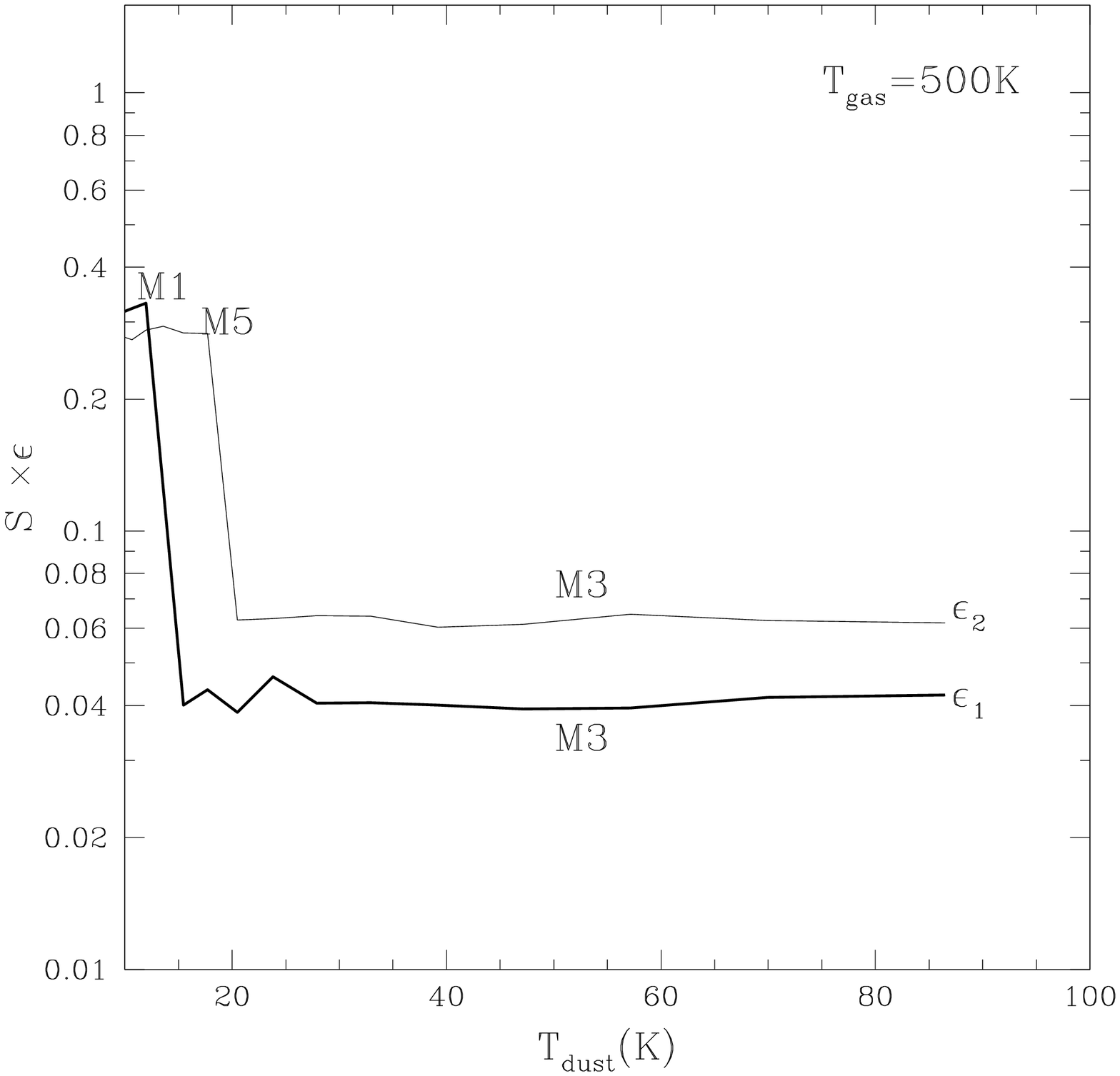}
\includegraphics[width=0.5\textwidth]{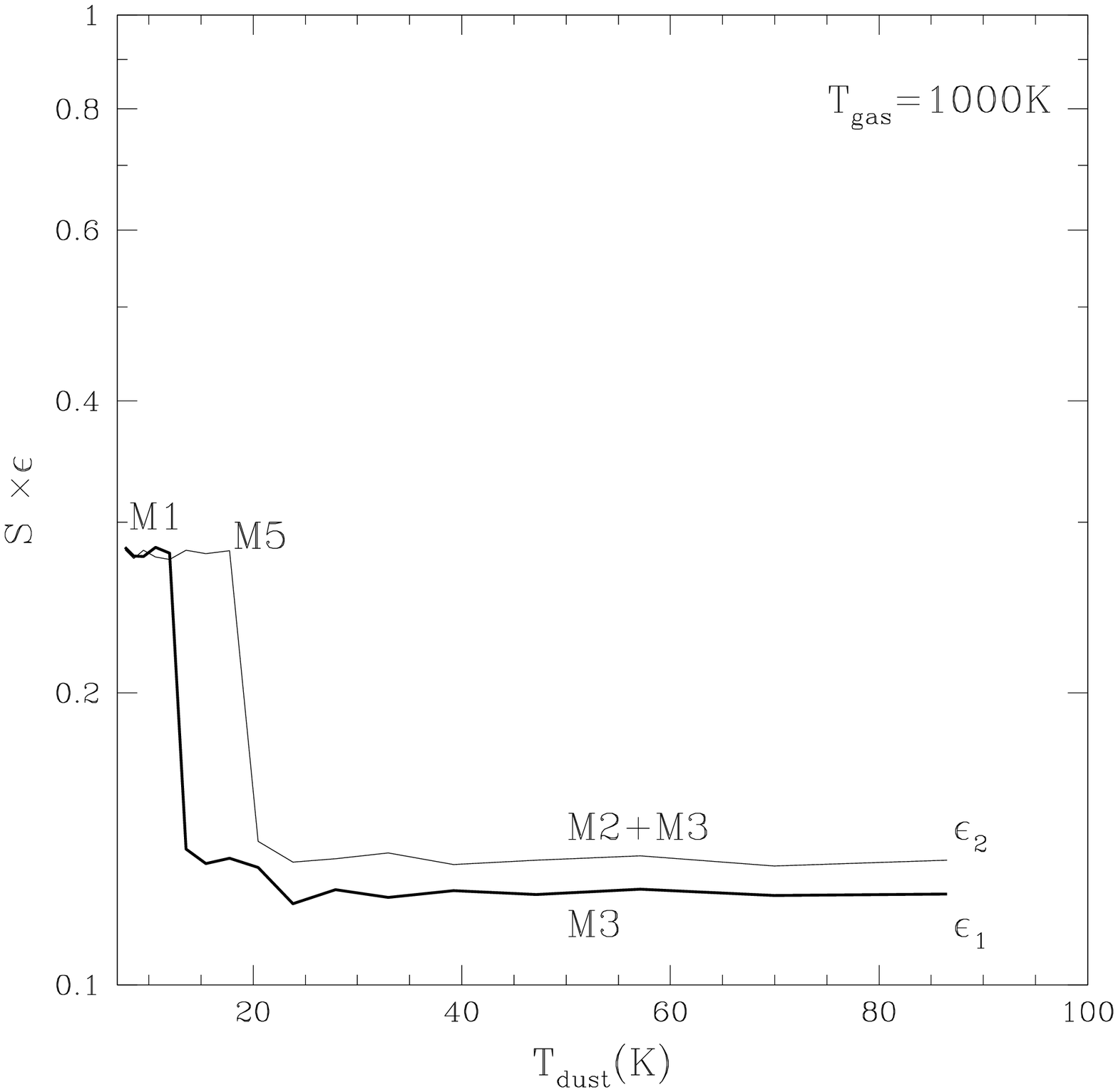}
\includegraphics[width=0.5\textwidth]{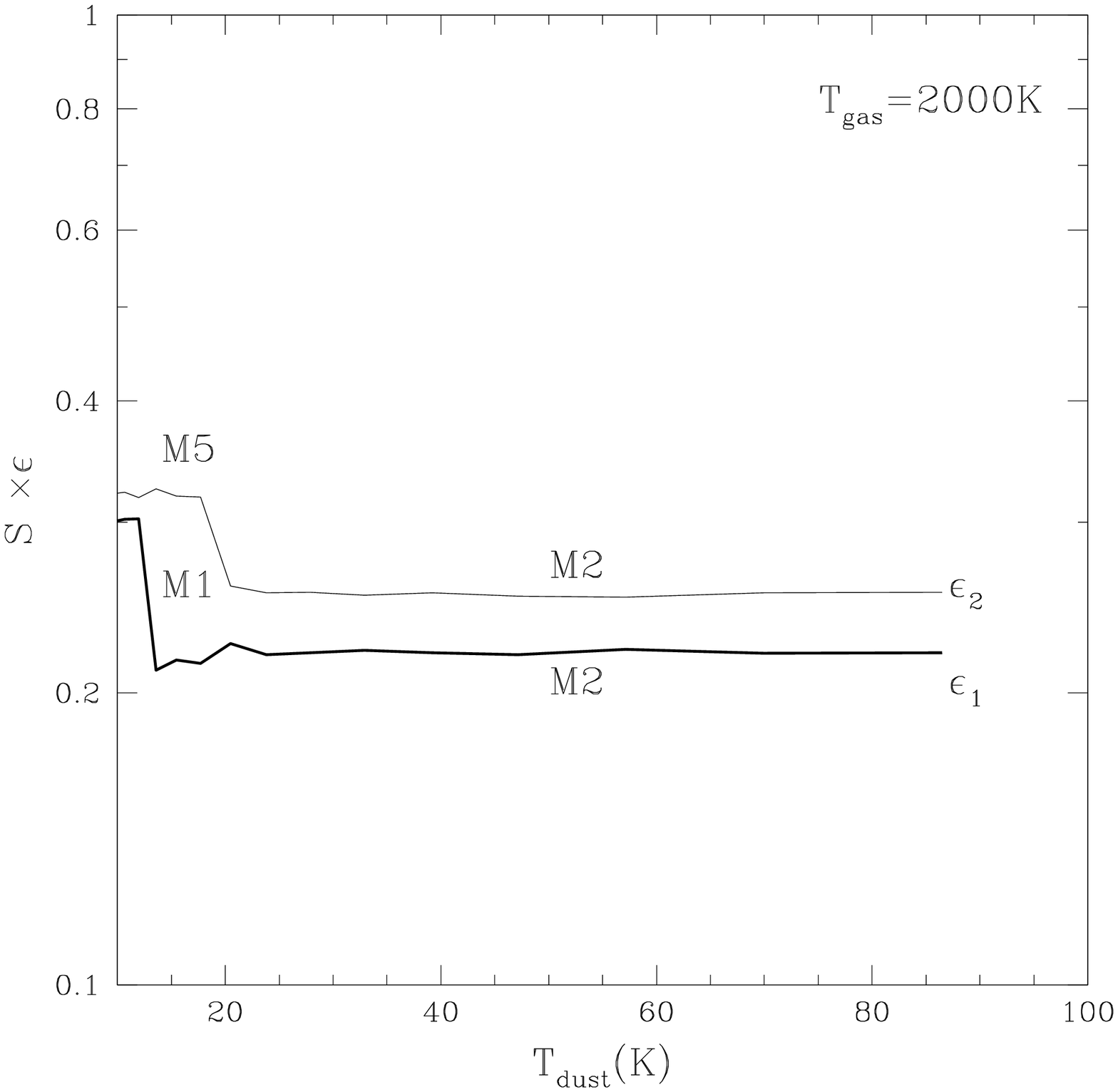}
\caption{H$_2$ formation efficiencies when squared barriers ($\epsilon_1$) and DFT barriers ($\epsilon_2$) are considered to enter in chemisorbed sites. The mechanisms important for the formation of H$_2$ are M1 (2 physisorbed atoms), M2 (direct chemisorption with H monomer) , M3 (direct chemisorption with H dimer) and M5 (physisorbed atom entering chemisorbed sites with H dimer) .}
\label{h2bar}
\end{figure}

\begin{figure}
\includegraphics[width=0.5\textwidth]{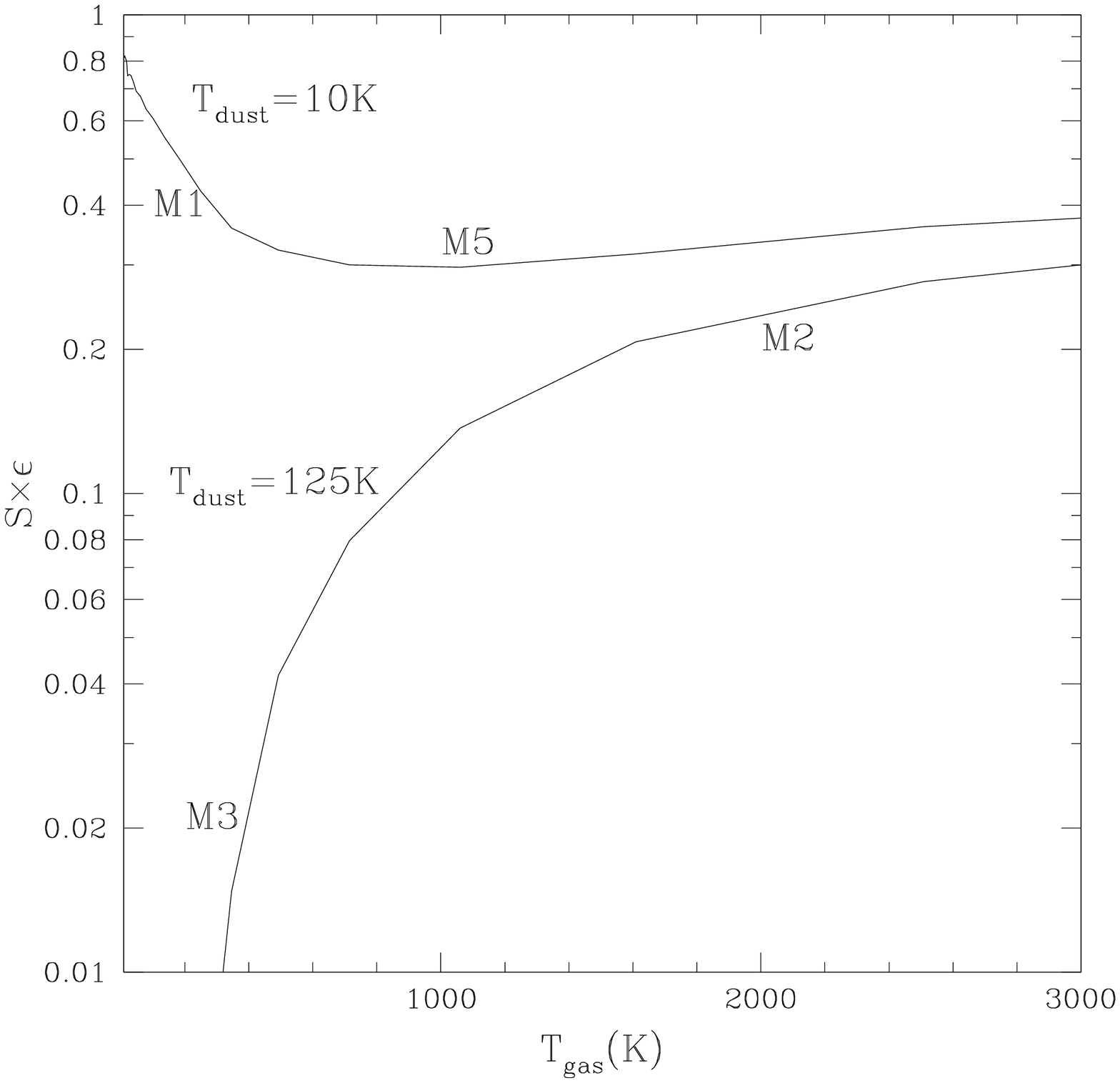}
\includegraphics[width=0.5\textwidth]{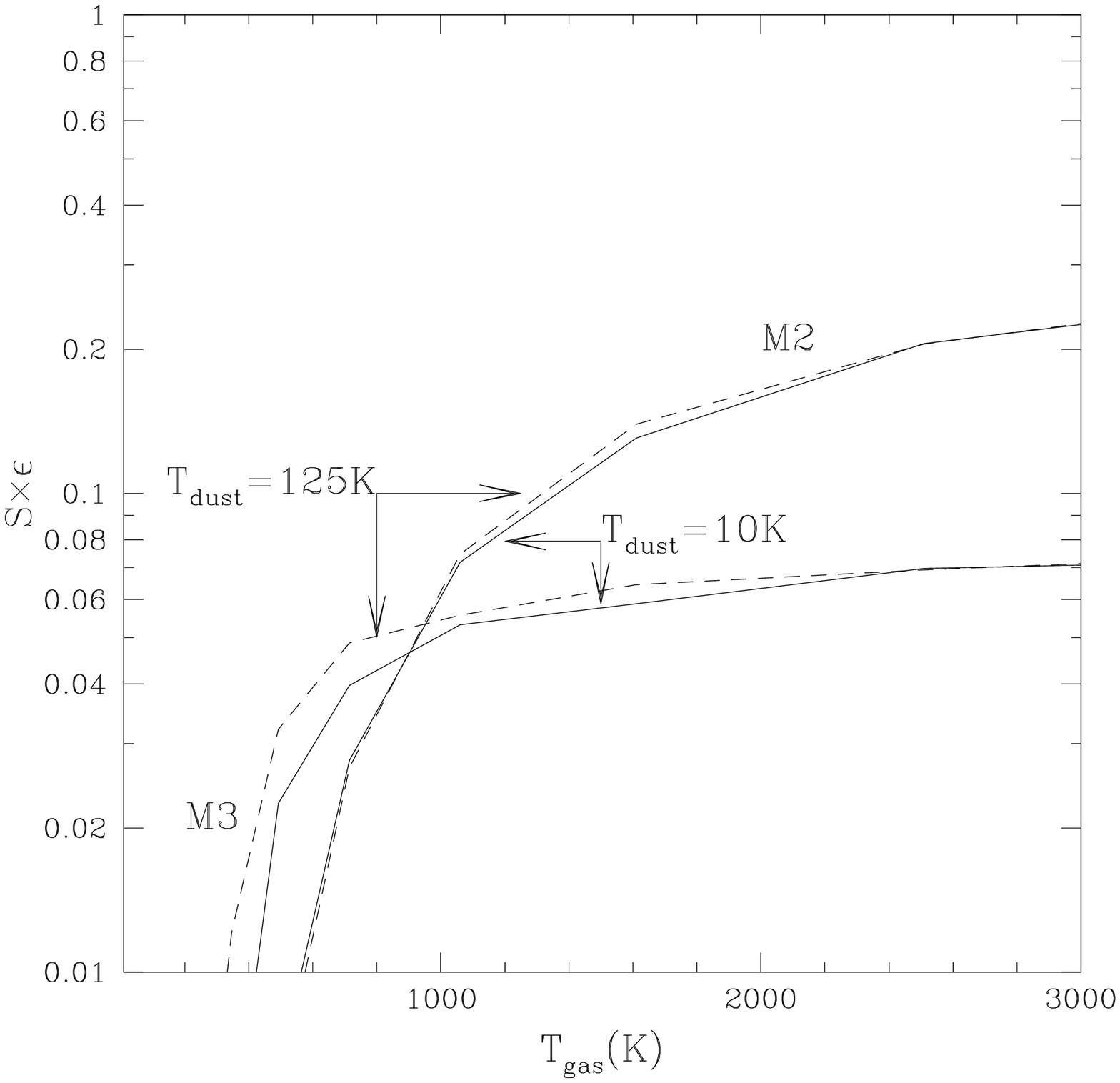}
\caption{H$_2$ formation efficiencies on 10~K and 125~K surfaces as a function of gas temperature, with the sticking in physisorbed sites S$_{phys}$ taken as in \cite{cuppen2010}. Left: total H$_2$ formation, governed by mechanisms M1 (2 physisorbed atoms) and M5 (physisorbed atom arriving in H dimer) for 10~K dust, and govern by mechanisms M3 (direct chemisorption with H dimer)  and M2 (direct chemisorption with H monomer). Right: Same as left but for H$_2$ formation involving only chemisorbed atoms. At low gas temperatures, H$_2$ is formed through mechanism M3 (direct chemisorption with H dimer).  For T$_{gas}$>1000~K, \hm\ is formed though mechamism M2 (direct chemisorption with H monomer).}
\label{phon1}
\end{figure}

\begin{figure}
\includegraphics[width=0.5\textwidth]{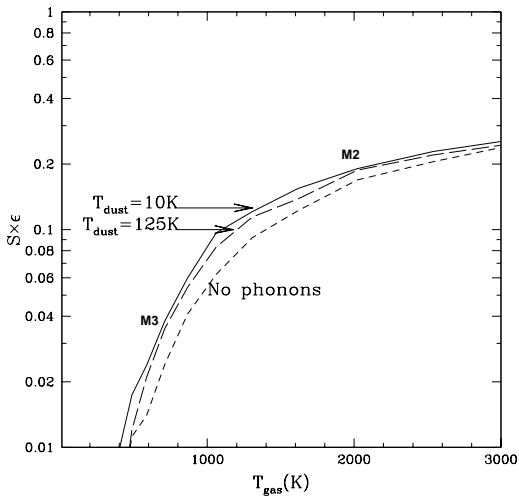}
\caption{Same as figure ~\ref{phon1} with the sticking in physisorbed sites S$_{phys}$ taken as in \cite{buch1989}.}
\label{phon2}
\end{figure}

\end{document}